\documentclass[10pt, conference, letterpaper]{IEEEtran}
\usepackage{cite}
\usepackage{amsmath,amssymb,amsfonts}
\usepackage{algorithmic,enumitem}
\usepackage{graphicx}
\usepackage{pdfpages}
\usepackage{textcomp}
\usepackage{xcolor}
\usepackage{multicol}
\usepackage{lipsum}
\usepackage{mwe}
\usepackage{subfigure}
\usepackage{subcaption, ragged2e}
\usepackage{diagbox}
\usepackage{slashbox}
\usepackage{dblfloatfix}
\usepackage{multicol}
\usepackage{multirow}

\begin{document}

\title{Locality Sensitive Hashing for Network Traffic Fingerprinting}

\author{
\IEEEauthorblockN{Nowfel Mashnoor, Jay Thom, Abdur Rouf, Shamik Sengupta, Batyr Charyyev}
\IEEEauthorblockA{Department of Computer Science and Engineering, University of Nevada, Reno, USA} 
\IEEEauthorblockA{
Email: nowfel@nevada.unr.edu, jthom@unr.edu, rouf@nevada.unr.edu, ssengupta@unr.edu, bcharyyev@unr.edu}
}

\IEEEoverridecommandlockouts
\IEEEpubid{\makebox[\columnwidth]{979-8-3503-4693-0/23/\$31.00~\copyright2023 IEEE\hfill} \hspace{\columnsep}\makebox[\columnwidth]{ }}

\maketitle
\IEEEpubidadjcol

\begin{abstract}
The advent of the Internet of Things (IoT) has brought forth additional intricacies and difficulties to computer networks. 
These gadgets are particularly susceptible to cyber-attacks because of their simplistic design. 
Therefore, it is crucial to recognise these devices inside a network for the purpose of network administration and to identify any harmful actions. 
Network traffic fingerprinting is a crucial technique for identifying devices and detecting anomalies. Currently, the predominant methods for this depend heavily on machine learning (ML). 
Nevertheless, machine learning (ML) methods need the selection of features, adjustment of hyperparameters, and retraining of models to attain optimal outcomes and provide resilience to concept drifts detected in a network. 
In this research, we suggest using locality-sensitive hashing (LSH) for network traffic fingerprinting as a solution to these difficulties. 
Our study focuses on examining several design options for the Nilsimsa LSH function. We then use this function to create unique fingerprints for network data, which may be used to identify devices.
We also compared it with ML-based traffic fingerprinting and observed that our method increases the accuracy of state-of-the-art by 12\% achieving around 94\% accuracy in identifying devices in a network.
\end{abstract}


\vspace{-1mm}
\section{Introduction}
Network traffic fingerprinting studies inferential methods that take the network traces of a group of devices as input and produce information about those devices, their users, and the environment~\cite{2018Mauro}. Network traffic fingerprinting is an important tool for network security and management as it enables to characterize traffic flows, identify the type of application such as streaming or web browsing, and match the traffic to the application or the device that generated it. Furthermore, it enables detecting and identifying anomalous flows and prevent malicious activities in a network. 
Traffic fingerprinting has been studied over a couple of decades focusing on operating system identification~\cite{2017Ahmet}, website detection~\cite{2019Meng}, and device fingerprinting~\cite{2014Sakthi}, etc. However, it regained the interest of researchers with the emergence of the Internet of Things (IoT) devices and the realization of smart systems.

Previous studies~\cite{2017Miettinen, 2019Ahmet, 2019Antonio} use network traffic fingerprinting for device identification. 
However, these traffic fingerprinting approaches rely on machine learning (ML) and require careful selection of features from network traffic data to prevent over-fitting/under-fitting the data and reduce computational overhead. 
There exist studies on feature selection for ML-based traffic fingerprinting that use optimization approaches such as Genetic Algorithms~\cite{2019Ahmet}, Grey Wolf Optimization~\cite{2019Amaal}, and Artificial Bee Colony~\cite{ghanem2020training}. 
ML-based approaches also require hyperparameter tuning and retraining the models if there exist newly available data, and concept drift in a network due to firmware and configuration update.
There exist studies on tuning and comparing ML models for network traffic fingerprinting~\cite{2019Eirini, 2019Antonio, 2019Bharat}. 
All these are computationally costly and add complexity to ML-based networking tools.

In our previous studies, we proposed using locality-sensitive hash (LSH) based network traffic fingerprinting for device identification~\cite{charyyevICC, charyyevIoTJ}, anomaly detection~\cite{charyyevGlobecom}, and user interaction classification~\cite{charyyevInfocom}. 
Locality-sensitive hash is different from cryptographic hashes and produces similar hash values for similar inputs. Thus we can use the hash similarity to identify devices by comparing the hash value of the network traffic data to hash values from different devices stored in a database. 
Since traffic fingerprinting with LSH uses the digest (i.e., hash) of the network traffic data we do not need to select and extract features from data. 
Also fingerprinting depends on hash similarity thus traffic fingerprinting with LSH does not have complex parameters that need to be tuned.
In \cite{charyyevICC, charyyevIoTJ} we proposed LSIF (Locality-Sensitive IoT Fingerprinting) to identify IoT devices with LSH-based network traffic fingerprinting. 
LSIF uses Nilsimsa~\cite{Damiani} as the underlying LSH function and achieves comparable results with state-of-the-art machine learning-based methods.
In this paper, we extend LSIF and propose LSIF-R (i.e., LSIF Reloaded), by exploring the alternative design of Nilsimsa. 
Specifically, we alter the implementation of Nilsimsa by trying different parameter configurations such as \textit{window size}, \textit{accumulator size}, etc., making the hash function more tunable. 
This will enable LSIF-R to be more robust to concept drift observed in a network, in addition to increasing the accuracy of device identification with traffic fingerprinting. 
We also compare LSIF-R with state-of-the-art IoTSentinel~\cite{2017Miettinen} and make a preliminary analysis of overhead and computational resource requirements.

Contributions of this paper are as follows:
\vspace{-2mm}
\begin{itemize}
    
    \item We extensively explore design alternatives for LSH function Nilsimsa for network traffic fingerprinting without feature extraction from the data, tuning the parameters of the model, and retrain the model.
    
    \item We propose a device identification system LSIF-R that uses LSH-based network traffic fingerprinting. LSIF-R is lightweight and robust to concept drifts.

    \item We compare LSIF-R with the state-of-the-art IoT device identification method IoTSentinel.
\end{itemize}


\section{Related Works}
In this section, we provide related works on network traffic fingerprinting, IoT device identification, and studies that use locality-sensitive hashing for various applications. 
We also briefly discuss how this work differs from our prior studies and existing related works.

\subsection{Network Traffic Fingerprinting}
Previous studies used network traffic fingerprinting for the operating system and device identification~\cite{2017Ahmet, 2017Miettinen, 2019Ahmet}, mobile application classification~\cite{van2020flowprint, trinh2020mobile}, and monitoring/anomaly detection~\cite{2018Meidan, 2018Yisroel}. 
Researchers have focused on different aspects of traffic fingerprinting such as feature selection~\cite{2019Ahmet, 2019Amaal, ghanem2020training}, model tuning~\cite{2019Eirini, 2019Antonio}, user privacy~\cite{zhang2022defeating, dietz2021browser}, fingerprinting in
NATed network, NAPT and VPN enabled networks~\cite{2019Shuaike, 2019Yair}, fingerprinting with background traffic~\cite{2019Samuel}, and fingerprinting on different states of the device~\cite{2019Ahmet, charyyevICC}. 

\subsection{IoT Device Identification}
Device identification is important for network management and monitoring~\cite{2019Ortiz}. 
Previous studies on IoT device identification with traffic fingerprinting use machine learning models~\cite{2019Eirini,
2017Miettinen, 2019Bharat} and signature-based methods~\cite{babun2020z, charyyevICC,charyyevIoTJ}.
There exist studies~\cite{2019Nesrine, 2019Franck} that use natural language processing models to identify devices with textual data extracted from mDNS, SSDP, DHCP, and HTTP protocols, and from DNS queries. 
There also exist studies that use one/multi-class classifiers and neural networks to identify devices with features extracted from wireless communication~\cite{2018Mustafizur,2019Antonio} and Bluetooth communication of the devices~\cite{2018Hidayet, 2018Tianbo}.

\subsection{Locality-Sensitive Hashing}
Previous studies use locality-sensitive hashing (LSH) for spam detection~\cite{marsono2012packet}, malware classification~\cite{friborg2019malware}, genome assembling~\cite{berlin2015assembling}, and content-based video retrieval~\cite{hu2005efficient}. Brian et al.~\cite{kulis2009kernelized} show that LSH enables accurate and fast performance with object classification, feature matching, and content-based retrieval. Muhammad et al. propose a technique for spam email fingerprinting at the packet level~\cite{marsono2012packet}. Locality-sensitive hashing was also used to recover the biometric information of individuals in a corporate environment to detect information leakage and similar events~\cite{alruban2018biometrically}. 
We also showed the effectiveness of LSH function Nilsimsa in identifying IoT devices~\cite{charyyevICC, charyyevIoTJ}, user interaction classification~\cite{charyyevInfocom}, and monitoring and detecting malicious activities in a network~\cite{charyyevGlobecom}.

Different from existing studies on network traffic fingerprinting~\cite{2017Ahmet, van2020flowprint, 2018Yisroel} and existing studies on device identification~\cite{2017Miettinen, 2019Ahmet} we propose a new approach with locality-sensitive hashing, which does not require feature selection/extraction from data and does not have complex hyperparameters that need to be tuned. 
Our study differs from existing studies on locality-sensitive hashing~\cite{marsono2012packet, friborg2019malware, hu2005efficient} by focusing on different utilization (i.e., device identification) of the hash function. Also compared to our previous studies~\cite{charyyevICC, charyyevInfocom, charyyevGlobecom} with Nilsimsa, our study differs by proposing an alternative implementation of Nilsimsa making the device identification system more flexible and at the same time robust to concept drifts observed in the network.

\begin{figure}[!b]
\includegraphics[width=0.8\columnwidth] {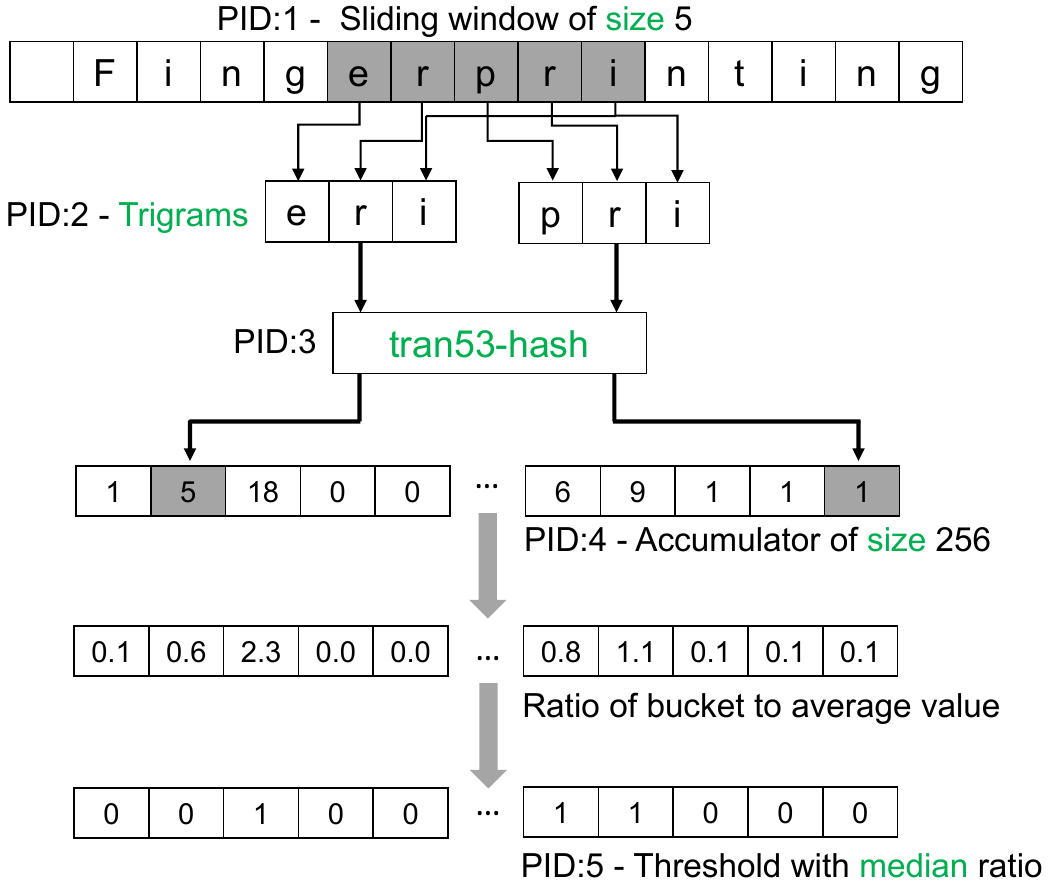}
\centering
\caption{Nilsimsa functionality}
\label{nilsimsafunction} 
\end{figure}

\section{Methodology}
This section will provide a concise overview of locality-sensitive hashing, with a special emphasis on Nilsimsa. 
We will provide comprehensive information about the alternative configuration of Nilsimsa and its practical execution. 
Next, we will provide comprehensive information on LSIF-R (Locality Sensitive IoT Fingerprinting-Reloaded) as well as the specifics of the dataset used for assessment purposes.

Nilsimsa is a locality-sensitive hash function that generates similar hash values for similar inputs. 
There are alternative LSH functions such as 
ssdeep, sdhash, and tlsh~\cite{gayoso2014state}. 
The functionality of Nilsimsa is presented in Figure~\ref{nilsimsafunction}. 
Nilsimsa uses a \texttt{sliding window} of size 5 characters that slide one character at a time. 
In each iteration it generates \texttt{trigrams} from data in the window and passes those trigrams to \texttt{tran54} hash function which generates value \texttt{i} between 0 and 255. 
Then \texttt{i-th} counter of the \texttt{accumulator} (i.e. array of integers of size 256) is increased by 1. After iterating over the entire input file the accumulator holds frequencies of the trigrams that have been found in the input. For each bucket in the accumulator, the ratio of the value in that bucket to the average values of the buckets will be assigned. 
Finally, if the \texttt{i-th} ratio is greater than the median of the ratios, the \texttt{i-th} bit of the Nilsimsa code will be set to 1 and it is set to 0, otherwise. The size of the hash generated by Nilsimsa is 32 bytes and the similarity score of two hashes ranges from -128 (i.e., completely different) to 128 (i.e., completely same). The similarity score is calculated by subtracting 128 from the number of similar bits in the hashes.

\subsection{LSIF-R}
The Nilsimsa hashes generated from different traffic flows of similar devices will have a high similarity score and a low score for hashes of traffic from different devices. 
Using this similarity we can identify devices if we have the signature (i.e., hash) of the traffic flow from devices in our database. 
We altered the design of Nilsimsa and made it more tunable. 
Specifically, we explored alternatives for parameters (i.e., window size, trigram, etc.). The set of parameters that we explored is shown in Figure~\ref{nilsimsafunction} with green color and preceded with their parameter ID (PID), and parameter values are presented in Table~\ref{parameterset}. 
The default values of Nilsimsa are shown with blue color on Table~\ref{parameterset}.
For \texttt{PID 1} (window size) we explored the values of 6, 7, 8, 9, and 10 alternative to default value 5. 
For \texttt{PID 2} we explored 3, 5, 7, 9 with the restriction of $n-grams < window size$, as we can not generate n-gram where \textit{n} is less than the window size. 
In terms of the underlying hash function (\texttt{PID 3}) and the threshold (\texttt{PID 5}) we used other classes of hashes such as SHA1-512, MD4-5, etc., and statistical values such as first quartile (Q1), third quartile (Q3), interquartile range (IQR), and standard deviation (std) as alternative threshold metric.
The default accumulator size in Nilsimsa is 256 and we used an accumulator size ranging from 16 to 1024 with an increment value of 16.

\begin{table}[!t]
\renewcommand{\arraystretch}{1}
\caption{Summary of the parameter set} 
\label{parameterset}
\resizebox{\columnwidth}{!}{
\begin{tabular}{l|c}
\textbf{Parameter ID} & \textbf{Values} \\ \hline
window size (PID 1)  & \textbf{\textcolor{blue}{5}}, 6, 7, 8, 9, 10 \\ \hline
n-grams (PID 2) & \textbf{\textcolor{blue}{3}}, 5, 7, 9 \\ \hline
\multirow{3}*{hash (PID 3)} & \textbf{\textcolor{blue}{TRAN53}}, SHA512, SHA384, SHA256, SHA224  \\
& SHA1, MD5, MD4, FNV164, FNV1A32, FNV1A64,  \\
& MMH3, CRC32, ADLER32, WHIRLPOOL \\
\hline
accumulator size (PID 4) & [16 : 1024], $\delta$ = 16 \quad | \quad \textbf{\textcolor{blue}{256}} \\ \hline
threshold (PID 5) & mean, \textbf{\textcolor{blue}{median}}, mode, IQR, Q1, Q3, std \\ \hline
\end{tabular}

}
\end{table}

\begin{figure}[!b]
\includegraphics[width=\columnwidth] {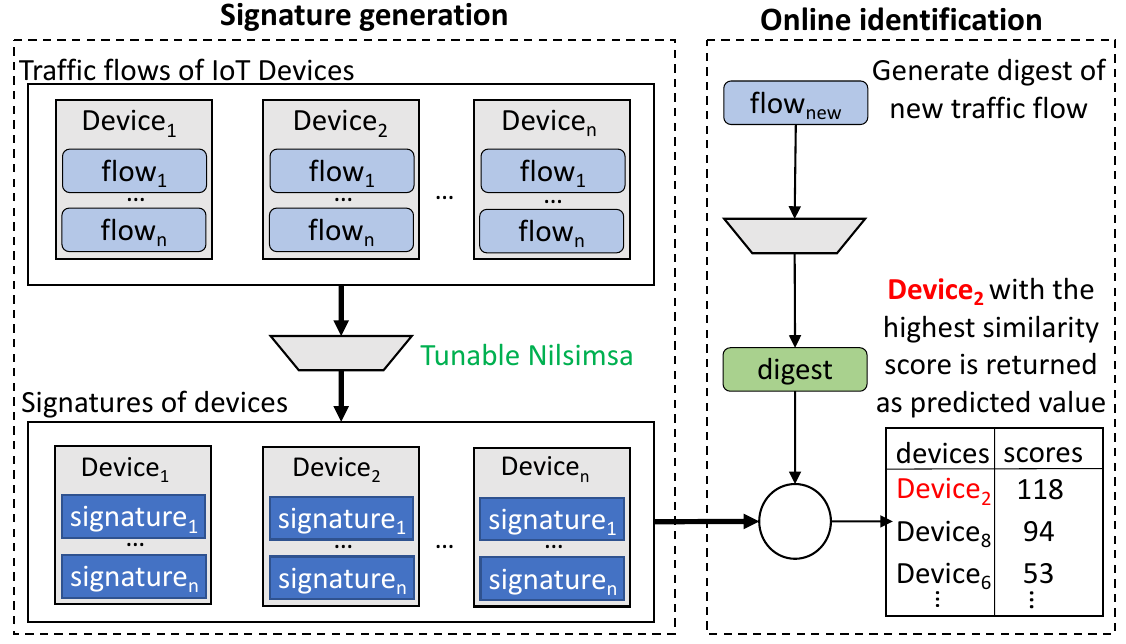}
\centering
\caption{System Design of LSIF-R}
\label{LSIFR} 
\end{figure}

In our previous studies~\cite{charyyevICC, charyyevIoTJ} we proposed a device identification system that uses Nilsimsa with its default parameters. 
In this paper, we propose IoT device identification system LSIF-R (Locality Sensitive IoT Fingerprinting Reloaded), that uses the tunable Nilsimsa described above. 
The design of the LSIF-R is presented in Figure~\ref{LSIFR}. 
First, LSIF-R computes a set of hash values of traffic flows generated by the devices and stores them as signatures in a database, with the device names as labels of signatures. 
These signatures are generated by tunable Nilsimsa.
Here traffic flow is a sequence of network packets sent and received by a particular IoT device in a certain time interval. 
Note that creating the initial signature database is performed only once, thus, it does not impose future overheads. 
When we try to identify the device we generate a hash of its traffic flow and compare it to signatures stored in a database. 
Then, the device with the highest average hash similarity score is selected as the predicted label for the device.
If the highest similarity score is less than a threshold value (i.e., min threshold for similarity score), then the device is labeled as unknown (i.e., the signature of the device does not exist in the database).
We can use average traffic flow similarity as the min similarity threshold value or we can employ a device-specific threshold based on the variance of the signatures of the device.

Compared to existing machine learning-based device identification systems~\cite{2019Ahmet, 2017Miettinen}, LSIF-R does not require feature selection and extraction from data, as it relies on the hash of the traffic flow. 
This provides some advantages over existing machine learning-based systems in terms of overhead as described in the next sections of our paper. 
Also it is simple to update LSIF-R if there is new available data. 
For instance, if we update the firmware of a particular device then its network characteristics might change and it needs to be reflected in the predictive model. If we want to integrate a new device into our system, again with ML-based device identification we need to re-train the model whereas with LSIF-R we only need to append signatures of the new device to the end of the signature database.
Thus, LSIF-R has clearly some benefits over machine learning based methods in terms of updating the model.


\subsection{Dataset and Metrics}~\label{dataset}
The dataset that we used for evaluation is from study~\cite{2017Miettinen}. 
It contains the network traffic of 23 IoT devices collected during the setup phase of the device. 
The dataset includes broad range of IoT devices ranging from sensors to home appliances. 
For each device, there exist 20 measurements. 
Since we use the hash of the traffic flow, we removed IP and MAC addresses associated with the devices to remove any bias that may occur in classifying the devices. 
We used 4-fold cross-validation for evaluations and performed 5 random experiments and presented the average. 
The evaluations were in terms of precision (i.e., $TP/(TP+FP)$), recall (i.e., $TP/(TP+FN)$), f1-score (i.e., $2/(1/precision + 1/recall)$), and accuracy (i. e., $(TP+TN)/(TP+FP+TN+FN)$) where TP, TN, FP, and FN stand for true positive, true negative, false positive and false negative, respectively.

\begin{figure}[!t]
\includegraphics[width=0.8\columnwidth] {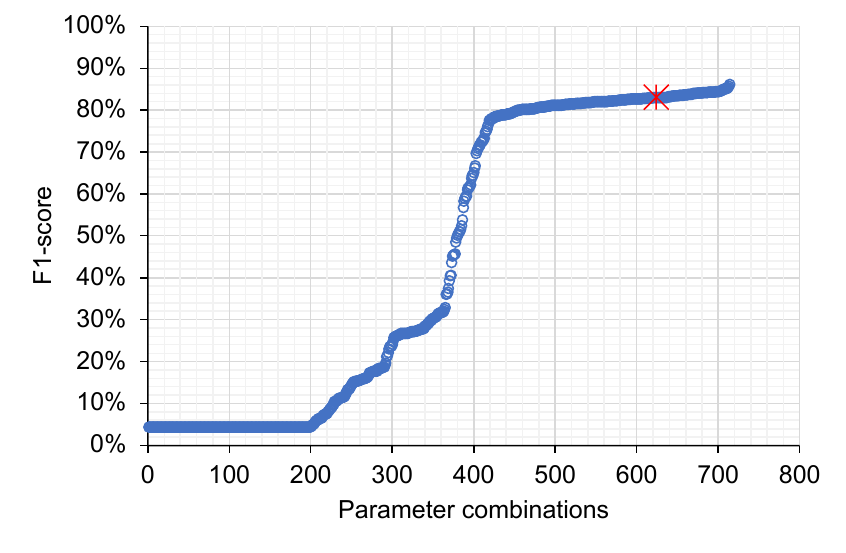}
\centering
\vspace{-2mm}
\caption{\small Device identification F1-score for default hash (Tran53) of Nilsimsa and trying different values for rest of parameters. The red star is F1-score for the default parameters of Nilsimsa.}
\label{tran_results}
\vspace{-4mm}
\end{figure}


\begin{figure}[!b]
\includegraphics[width=\columnwidth] {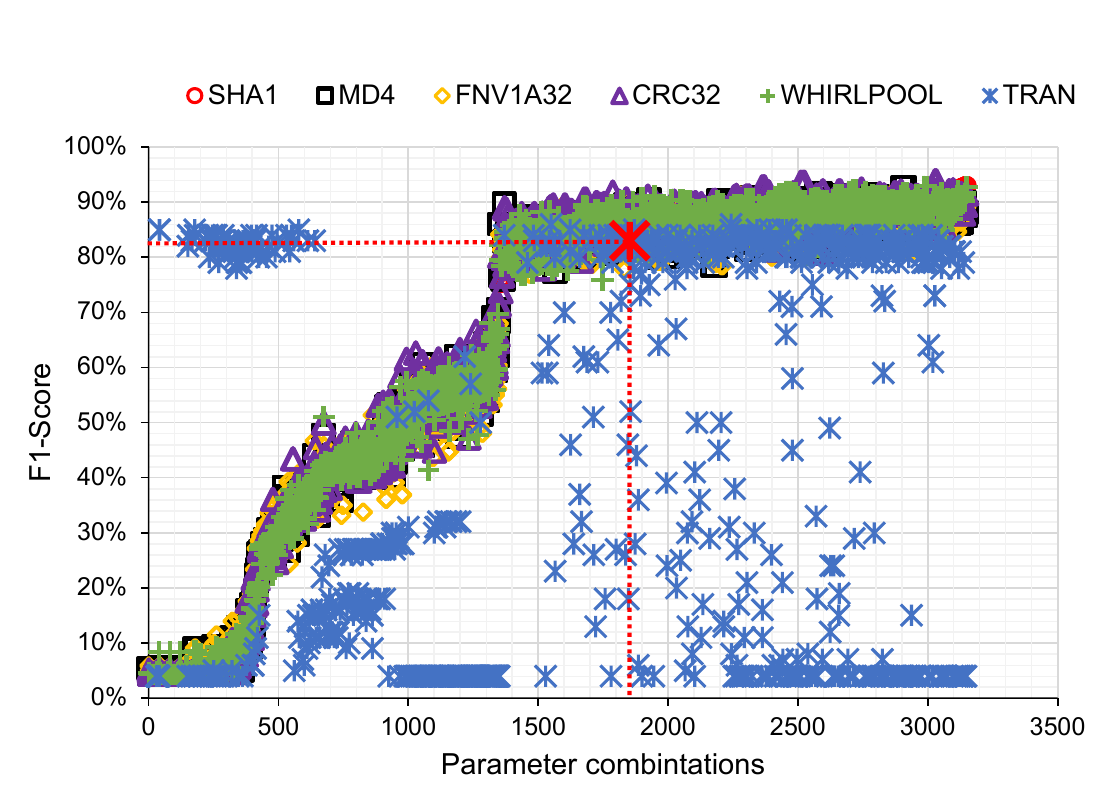}
\centering
\caption{F1-score for different classes of hash functions.}
\label{tran_others} 
\end{figure}

\section{Evaluation and Results}
In this section, we provide evaluation results of the proposed device identification method LSIF-R in identifying 23 IoT devices described in Section~\ref{dataset}.
We first evaluate the LSIF-R by trying different parameters from Table~\ref{parameterset}. 
Then we focus on if there is any correlation between the hyperparameter set and accuracy. 
Finally, we compare LSIF-R with the existing state-of-the-art method IoTSentinel~\cite{2017Miettinen}.

\subsection{Design alternatives of Nilsimsa}~\label{s1}
\begin{figure}[!b]
\vspace{-2mm}
\includegraphics[width=\columnwidth] {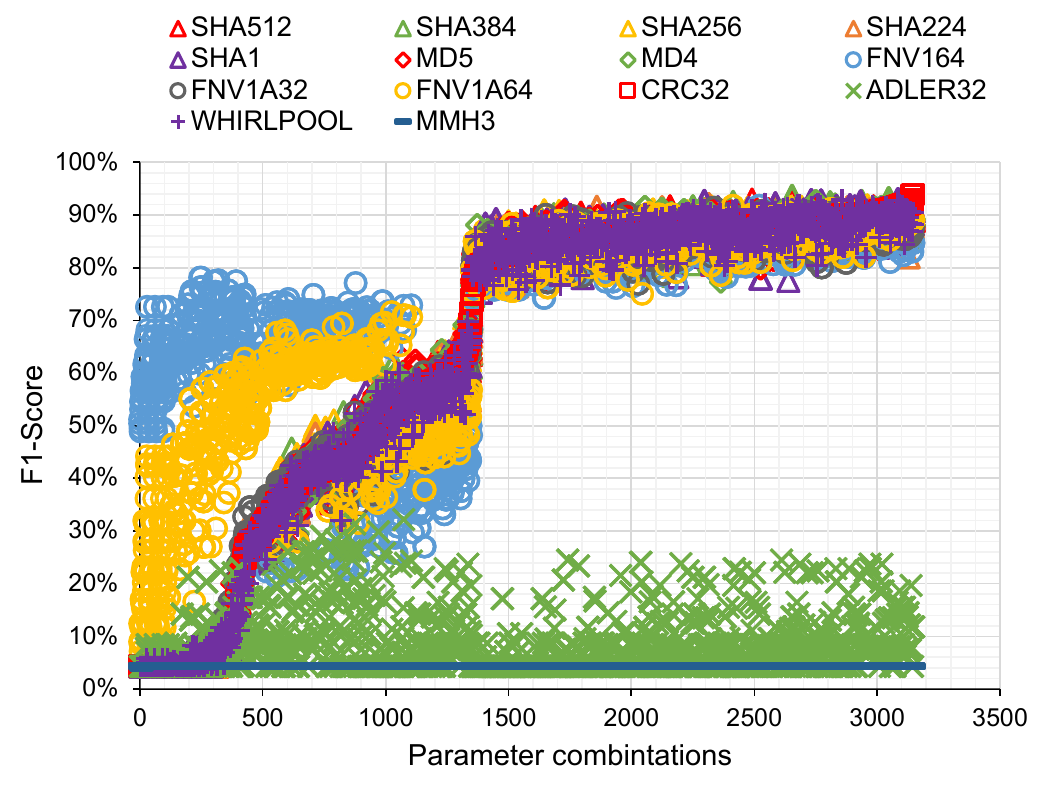}
\centering
\caption{F1-score of LSIF-R with other hash functions.}
\label{allhashes} 
\end{figure}

First, we started the evaluation by trying different hyperparameters from Table~\ref{parameterset} while keeping the underlying hash (Tran53) function of Nilsimsa unchanged. 
Figure~\ref{tran_results} presents results in terms of the F1-score. 
When Tran53 is stable, we can increase the average accuracy by 3\% from the accuracy of 83\% achieved with default parameters of Nilsimsa (shown with a red star in Figure~\ref{tran_results}). 
The change in accuracy increases even more when we try different hash values instead of Tran53. 
Figure~\ref{tran_others} shows the results when we try different hash functions from other hash classes such as SHA, MD, FNV, etc. 
We can achieve an average F1-score above 90\% with other classes of hash functions. 
For instance SHA1 achieves F1-score of 93\%, MD4 achieves 93\%, and FNV164 and CRC32 achieves 92\% and 94\% of F1-scores respectively. 
These values show that we can increase the accuracy of LSIF-R by 10\% by trying different hyperparameters for Nilsimsa and changing its design and implementation. 
While a majority of functions achieve similar results, not all hash functions achieve good results.
For instance in Figure~\ref{allhashes} we can see that hash functions MMH3 and ADLER32 perform very poorly in terms of device identification for all parameter combinations. 
However, we can also see that hash functions FNV164 and FNVA164 achieve slightly better performance for parameters between 0-1000 compared to the rest of the hash functions. 
This suggests that we can use the ensemble of FNV164 and FNVA164 with other hash functions to increase accuracy for parameters between 0-1000.
This might be useful if parameters between 0-1000 are more lightweight compared rest of the parameter combinations. 
For instance assume the $900^{th}$ parameter has \textit{window size} and \textit{n-gram} equal to 5 and 3 respectively and assume these values for parameter $2000^{th}$ are 9 and 3. 
In this case, parameter $900^{th}$ will be more lightweight than $2000^{th}$ because $Combination(5,3) < Combination(9,3)$, and $Combination(5,3)$ indicates how many trigrams will be generated with window size equal to 5 and n-gram equal to 3.
LSIF-R achieves higher accuracy on the parameter $2000^{th}$ with single hash functions. 
However, if it can achieve comparable performance with a combination of multiple hash functions at parameter $900^{th}$ then ensembling the hash functions needs to be considered as using the parameter combination $900^{th}$ is more lightweight than parameter $2000^{th}$. 
Overall we observed that if we tune the other parameters of Nilsimsa and use the hash functions such as SHA, MD instead of default TRAN53 the accuracy of Nilsimsa-based LSIF-R reaches up to 94\%. 
From alternative hash functions for Nilsimsa, we observed that CRC32 achieves the best result thus in the rest of the paper our evaluations will be based on CRC32.

\begin{figure}[!t]
\includegraphics[width=0.9\columnwidth] {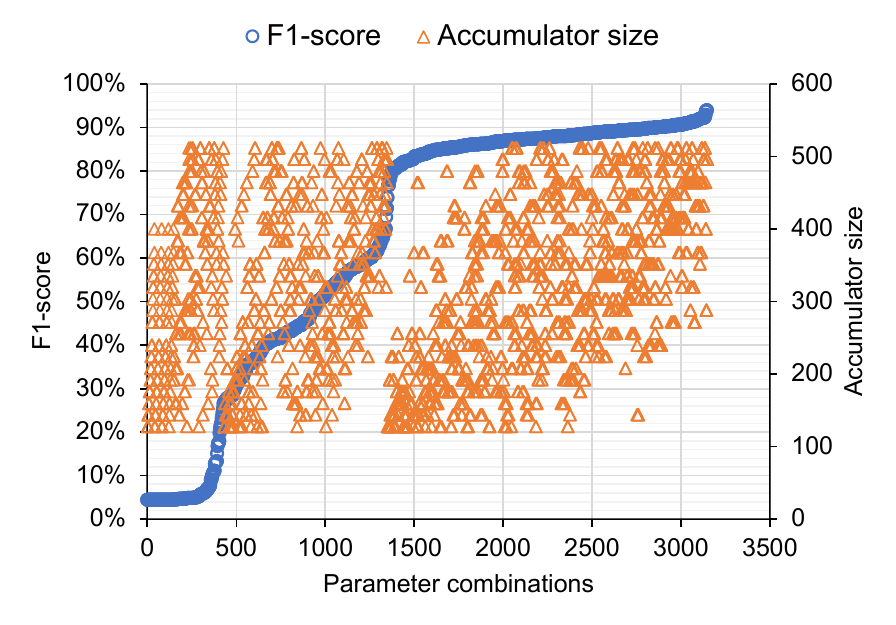}
\centering
\vspace{-2mm}
\caption{Correlation of individual parameter with accuracy.}
\vspace{-4mm}
\label{correlation} 
\end{figure}

\subsection{Correlation with hyperparameters}
In Figure~\ref{allhashes}, if we remove MMH3, ADLER32, FNV164, and FNVA164 hashes, we can see that the rest of the hashes have very similar behavior. 
We can also see that for parameters above $1500^{th}$ the accuracy of LSIF-R reaches around 90\% with all hashes. 
Thus, it is important to analyze which parameters achieve high accuracy and if there is a correlation between accuracy and any specific parameter. 
Figure~\ref{correlation} shows the correlation between the accuracy of LSIF-R and \textit{accumulator size} one of the hyperparameters of Nilsimsa from Table~\ref{parameterset}. 
We can see that there is no correlation between accuracy and accumulator size. 
We observed similar behavior for other parameters such as \textit{window size}, \textit{n-grams}, and \textit{threshold type}, which are not shown due to space constraints. 
While there is no correlation between accuracy and individual parameters, there might exist some correlation between accuracy and a combination of multiple parameters. 
These analyzes are important as the overhead of Nilsimsa might change depending on the values of hyperparameters. 
As we discussed in subsection~\ref{s1} the parameter pair (window size=5, n-gram=3) will have low overhead compared to the parameter pair (window size=9, n-gram=3). 
Overall, we did not observe any correlation between individual parameters and accuracy however exploring the overhead for different parameter combinations and the trade-off between accuracy and imposed overhead from parameter combinations need to be explored in the future. 

\subsection{Comparison with related works}
In this subsection we compared LSIF-R with the state-of-the-art machine learning-based IoT device identification system IoTSentinel~\cite{2017Miettinen}. 
IoTSentinel works as follows: it extracts the 23 network traffic features from the first 12 packets of the device and generates a feature matrix and feeds it to the machine learning model Random Forest. 
The features include the existence of Link/Network/Transport layer protocols, data related to IP headers, packet count, etc. 
A Random Forest model is trained for each device and it makes binary classification (if traffic belongs to that device or not). 
If traffic flow belongs to multiple or none of the devices as a result of random forest classifications then edit distance discrimination is used to identify the device. 
Different from the IoTSentinel, LSIF-R is simple and it is easy to update the model if there is new available data. 
For instance, if we need to update the device identification system due to a firmware update of that device or if we want to integrate the new device into the system we will need to update the whole underlying machine learning model of the system. 
Whereas this is not the case with LSIF-R where we only need to update the signatures of a particular device or append the signatures of a new device to the signature database. 
Also, LSIF-R does not require feature selection as it only takes the hash of the traffic flow. 
The feature selection requires expert knowledge and it is computationally expensive. 
For instance, IoTSentinel uses the first 23 packets to extract the features from data. 
The features include the existence of a particular protocol in data or statistics about source/destination port numbers and IP numbers. 
Since these features are computed over the first 23 packets we need to store the information related to 23 packets and wait till those packets are generated, making feature selection costly in terms of storage and time. 
This cost might be significant especially if IoTSentinel is deployed on embedded devices with limited computational resources. 
However, this is not the case with LSIF-R as it uses the hash of the traffic flow and the hash can be generated as network traffic data is generated. 
For instance in our previous study~\cite{charyyevIoTJ} we observed that the processing time of Nilsimsa with default configuration is $3.3\times10^5$ bytes/s which is significantly higher than the traffic rate of IoT devices used in our evaluation (WeMoLink has the highest rate of the traffic 992 bytes/s).
Since the processing speed of Nilsimsa is significantly higher than the traffic rate of devices we can generate the hash of the traffic as it comes with negligible overhead.
These suggest that LSIF-R is more lightweight than machine learning-based device identification systems in terms of feature selection/extraction and updating the model.

\begin{figure}[!b]
\vspace{-4mm}
\includegraphics[width=0.9\columnwidth] {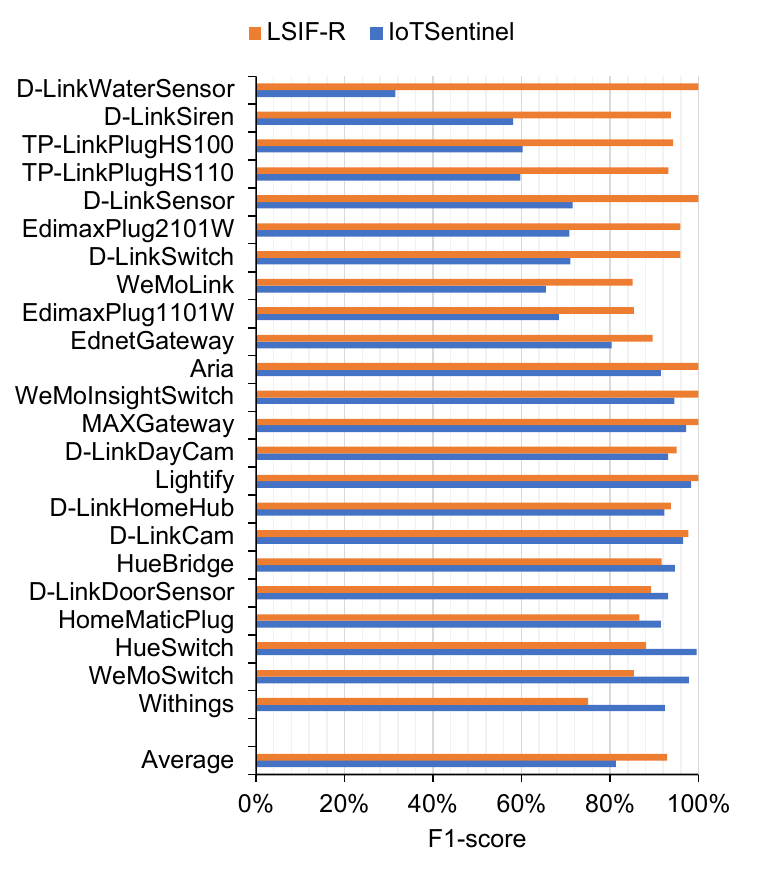}
\centering
\vspace{-3mm}
\caption{Comparison of LSIF-R and IoTSentinel in identifying 23 devices. The average and individual results are presented.}
\label{comparison} 
\end{figure}

In terms of accuracy, IoTSentinel achieves average precision, recall, and f1-score of 84\%, 82\%, and 81\% respectively and these values for LSIF-R are 94\%, 93\%, and 93\%. 
Figure~\ref{comparison} presents the average performance comparison of IoTSentinel and LSIF-R and the performance comparison for individual devices. 
We can see that LSIF-R achieves better accuracy in 17 devices out of 23, and in terms of average accuracy LSIF-R has an average F1-score 12\% higher than the IoTSentinel. 
These findings show that LSIF-R is more lightweight and achieves higher accuracy compared to the state-of-the-art machine learning-based device identification system IoTSentinel.

\section{Conclusion}
Network traffic fingerprinting is an important tool for network management and monitoring. In this paper, we proposed network traffic fingerprinting with locality-sensitive hashing.
Specifically we explored design alternatives for the hash function Nilsimsa and proposed LSIF-R a device identification method that uses Nilsimsa-based network traffic fingerprinting. 
Since LSIF-R depends on locality sensitive hashing it does not require feature selection and extraction from traffic data.
Also by design LSIF-R does not have complex parameters that need to be tuned and it is relatively straightforward to update it with new measurements. 
We compared the LSIF-R with IoTSentinel which uses machine learning-based network traffic fingerprinting to identify devices. 
We observed that LSIF-R has higher accuracy (by 12\%) than IoTSentinel and achieves 94\% accuracy in identifying 23 IoT devices. 
In the future, we want to explore the computational overhead of different parameter combinations of Nilsimsa on LSIF-R and explore the trade-off between accuracy and imposed overhead. 
We also want to explore self-tuning properties for LSIF-R to mitigate the impact of concept drift observed in a network.

\bibliographystyle{IEEEtran}
\bibliography{references}

\end{document}